\documentclass[final,1p,times]{elsarticle} % do not change
\usepackage{graphicx} % do not change
\usepackage{amssymb} % do not change
\usepackage{amsthm} % do not change
\journal{Nuclear Physics A} % do not change
\begin{document} % do not change

\begin{frontmatter} % do not change

%% QM09Author: please enter your  
%% Title, author and address info here; please do not use footnotes

% Your Title - please modify
\title{Bulk Properties in Au+Au Collisions at $\sqrt{s_{NN}}$ = 9.2 GeV in STAR Experiment at RHIC}

% Principle author, and co-authors - please modify
%\author{Irakli Garishvili$^{a}$ for the PHENIX collaboration}
\author{Lokesh Kumar$^{a}$ for the STAR collaboration}

% Address - please modify
% note that if you have authors from several institutions, we recommend
% labelling these [a], [b], [c] etc.
\address[a]{Panjab University Chandigarh, % label [a]
Sec. - 14,
Chandigarh, 160014, INDIA}

\begin{abstract} % do not change
%% Text of abstract goes here - please modify
One of the primary goals of high-energy heavy-ion collisions is to
establish the QCD phase diagram and search for possible phase
boundaries. The planned RHIC energy scan program will explore this
exciting physics topic using heavy-ion collisions at various
center of mass energies. The first test run with Au+Au collisions at
$\sqrt{s_{NN}}$ = 9.2 GeV took place in early 2008. 
We present the results on identified particle ratios, azimuthal 
anisotropy parameters ($v_{1}$ and $v_{2}$) and HBT
%and {\color{red} $\pi$ interferometery} 
at midrapidity %for Au+Au collisions at $\sqrt{s_{NN}}$ = 9.2 GeV.
using data from this run.
These results are compared to data for both lower
and higher center of mass energies at the AGS, SPS and RHIC.
These new data demonstrate the capabilities of the STAR detector for exploring 
the QCD phase diagram.
\end{abstract} % do not change

\end{frontmatter} % do not change

%% QM09: we keep linenumbers at least for initial version
%\linenumbers % do not change

%% start of main text - please modify. Below is a sub-set (single section) 
%% of an earlier proceedings that show how one can handle references 
%% and figures etc.
%%\section{}\label{}

%%%%%%%%%%%%%% copied %%%%%%%%%%%%%%%%%%%
\section{Introduction: }
Searches for the Quantum Chromodynamics (QCD) critical point and for the location of
the phase boundaries in the QCD phase diagram have been of great interest in
high-energy heavy-ion collisions, both 
theoretically and experimentally.
The QCD phase diagram is usually plotted as 
%baryonic chemical potential ($\mu_{B}$) vs. temperature (T).
temperature (T) vs. baryonic chemical potential ($\mu_{B}$).
%QCD critical point in the phase diagram represents a point where the
A critical point in the QCD phase diagram is the location where
first order phase transition ends. 
Phase boundaries in the diagram distinguish the hadronic phase from the Quark Gluon Plasma
(QGP) phase. 
Lattice calculations suggest that near $\mu_{B}$ $\sim$ 0, a smooth crossover
occurs between %the hadronic phase and the Quark Gluon Plasma (QGP) phase.
%Phase boundaries in the diagram distinguish the hadronic phase from the QGP phase. 
these two phases.
To search for the QCD critical point and 
to explore the phase diagram, we need to vary the $\mu_{B}$ and temperature.
These can be varied by altering the center of mass energy
and are deduced from the spectra and ratios of produced particles.
%on comparing with the model calculations. 
%In view of these,
%There is a proposal from STAR to start a new program called ``Critical Point Search'', in
There is a proposal at RHIC to start a new program called ``Critical Point Search'', in
which $\sqrt{s_{NN}}$ will be varied in order to explore the QCD phase diagram.
In addition to the critical point search, the STAR experiment would like to locate the 
$\sqrt{s_{NN}}$ where many interesting observations at top RHIC energies, 
such as number of constituent quarks (NCQ) scaling of $v_{2}$~\cite{qscale},
high transverse momentum ($p_{T}$) hadron suppression in A+A collisions relative to p+p
collisions~\cite{suppression} and the ridge formation~\cite{ridge},
will ``disappear'' or ``switch off''.
%As a first step of ``Critical Point Search'' program, a test run was conducted at
As a first step of this program, a test run was conducted at
RHIC in 2008 by colliding Au ions at $\sqrt{s_{NN}} = 9.2$ GeV. 
This 
short test run yielded $\sim$3000 good events and the results presented in this
paper are based on these data.
%%%%%%%%%%%%%% Adding event selection %%%%%%%%%%%%%%
These number of events were obtained by selecting the $z$-position of the event vertex 
within 75 cm of the nominal collision point.
%%%%%%%%%%%%%% Adding event selection %%%%%%%%%%%%%%

\section{Results} 
\subsection{Energy dependence of particle ratios}
\begin{figure}
\begin{center}
\vspace{-0.3cm}
\includegraphics[height=10cm]{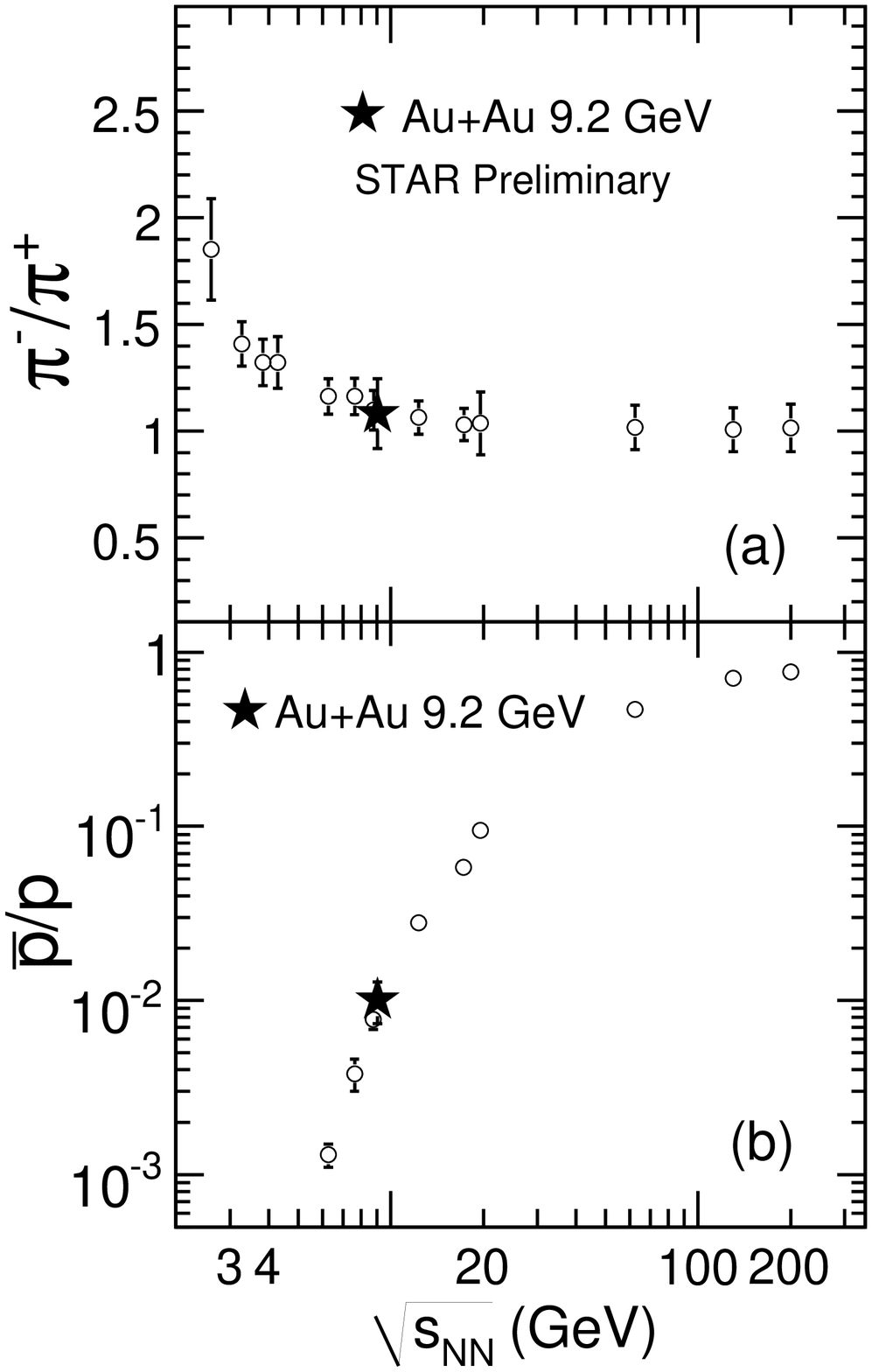}
\includegraphics[height=10cm]{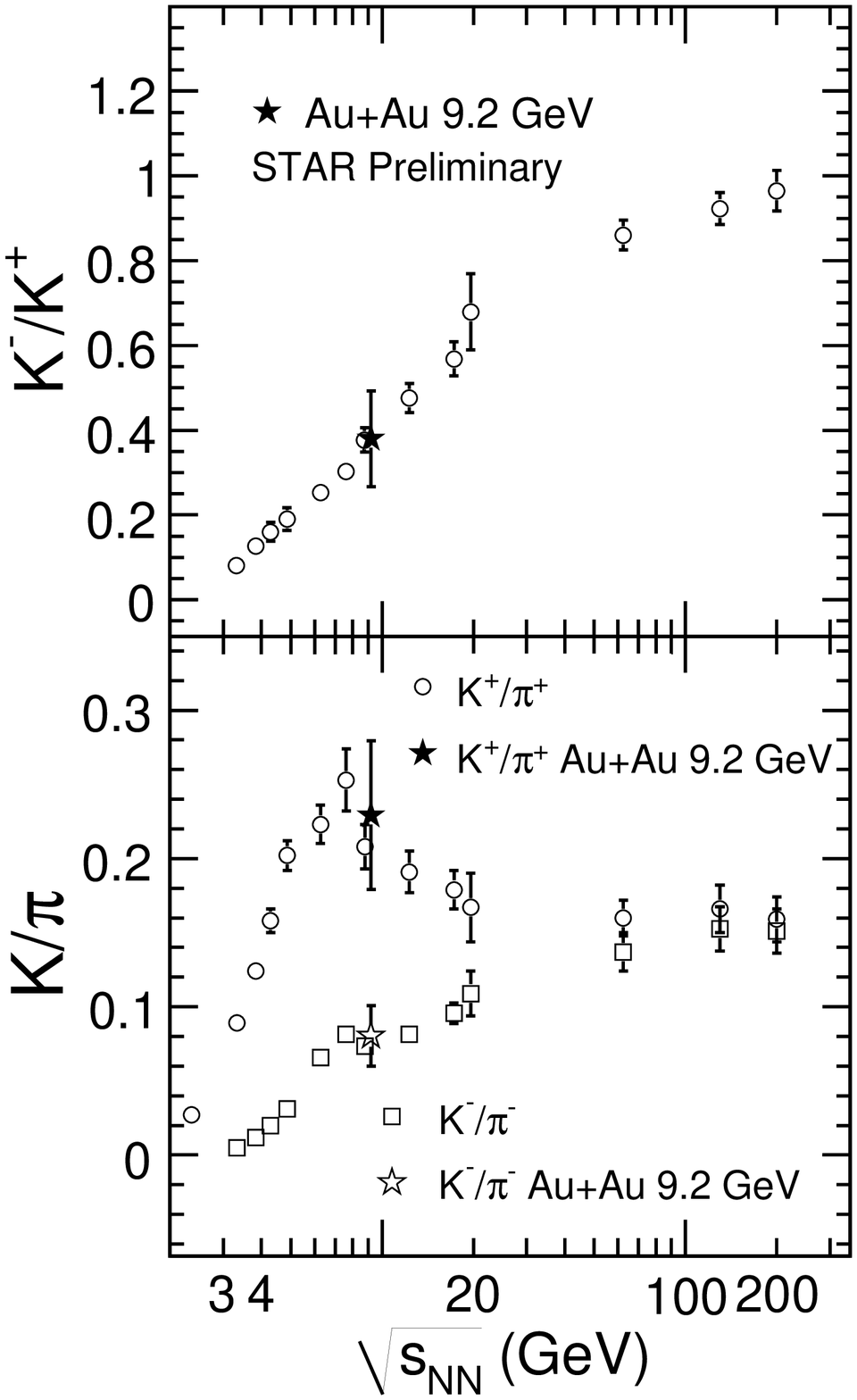}
\vspace{-0.5cm}
\caption{\label{fig:epsart} Left panel: (a) $\pi^{-}$/$\pi^{+}$ and (b) $\bar{p}$/$p$,
plotted as a function of $\sqrt{s_{NN}}$. Right panel: (a) $K^{-}$/$K^{+}$ and (b) $K$/$\pi$,
plotted as a function of $\sqrt{s_{NN}}$. 
Results from 0--10\% central Au+Au collisions at 9.2 GeV (solid stars) are compared with those from 
AGS~\cite{ags}, SPS~\cite{sps} and RHIC~\cite{STARPID} (open symbols).
Errors are statistical and systematic added in quadrature. See text for details.
}
\end{center}
\end{figure}

\begin{figure}
\begin{center}
\vspace{-0.3cm}
\includegraphics[height=13.5pc]{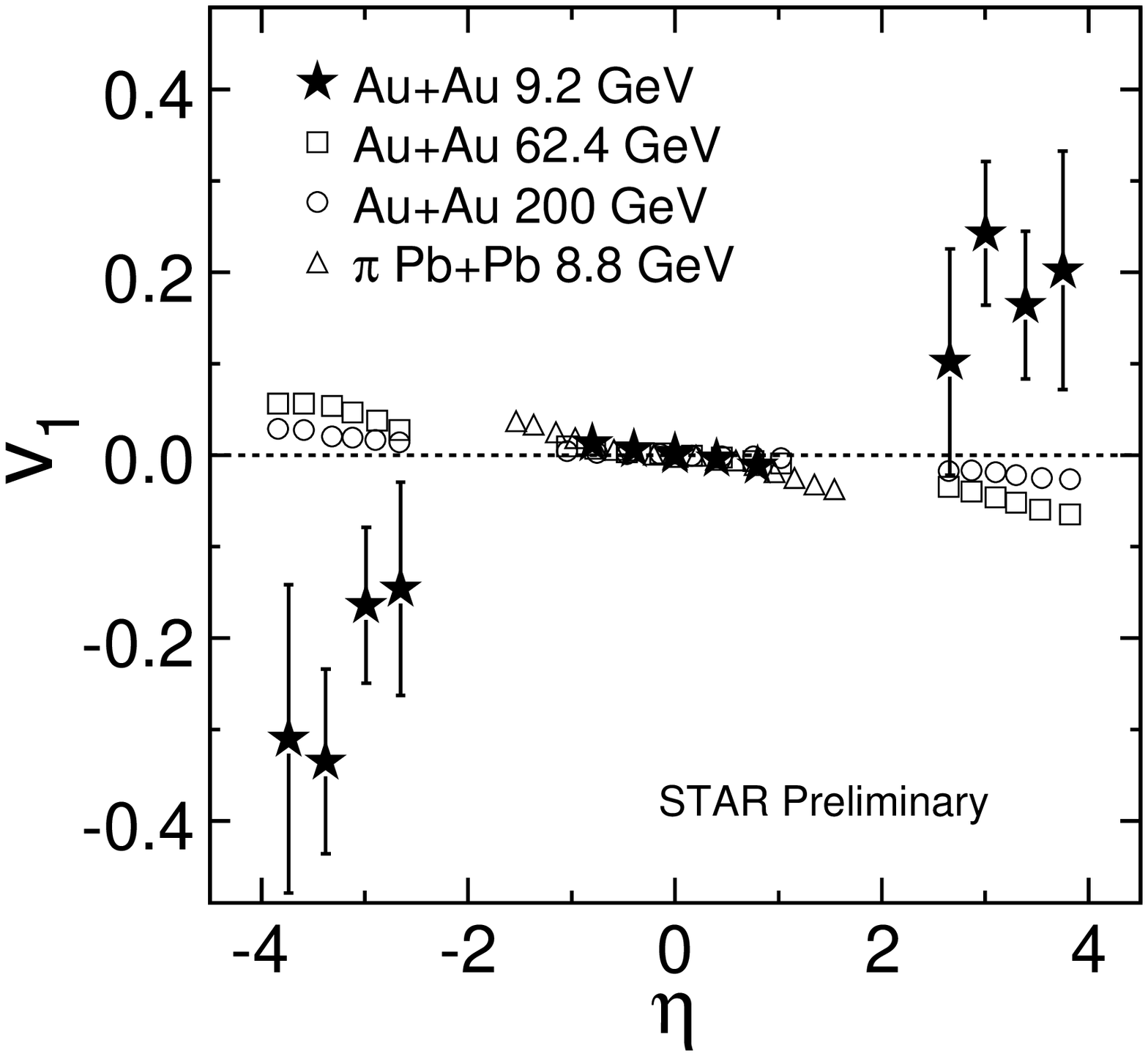}     
\includegraphics[height=13.5pc]{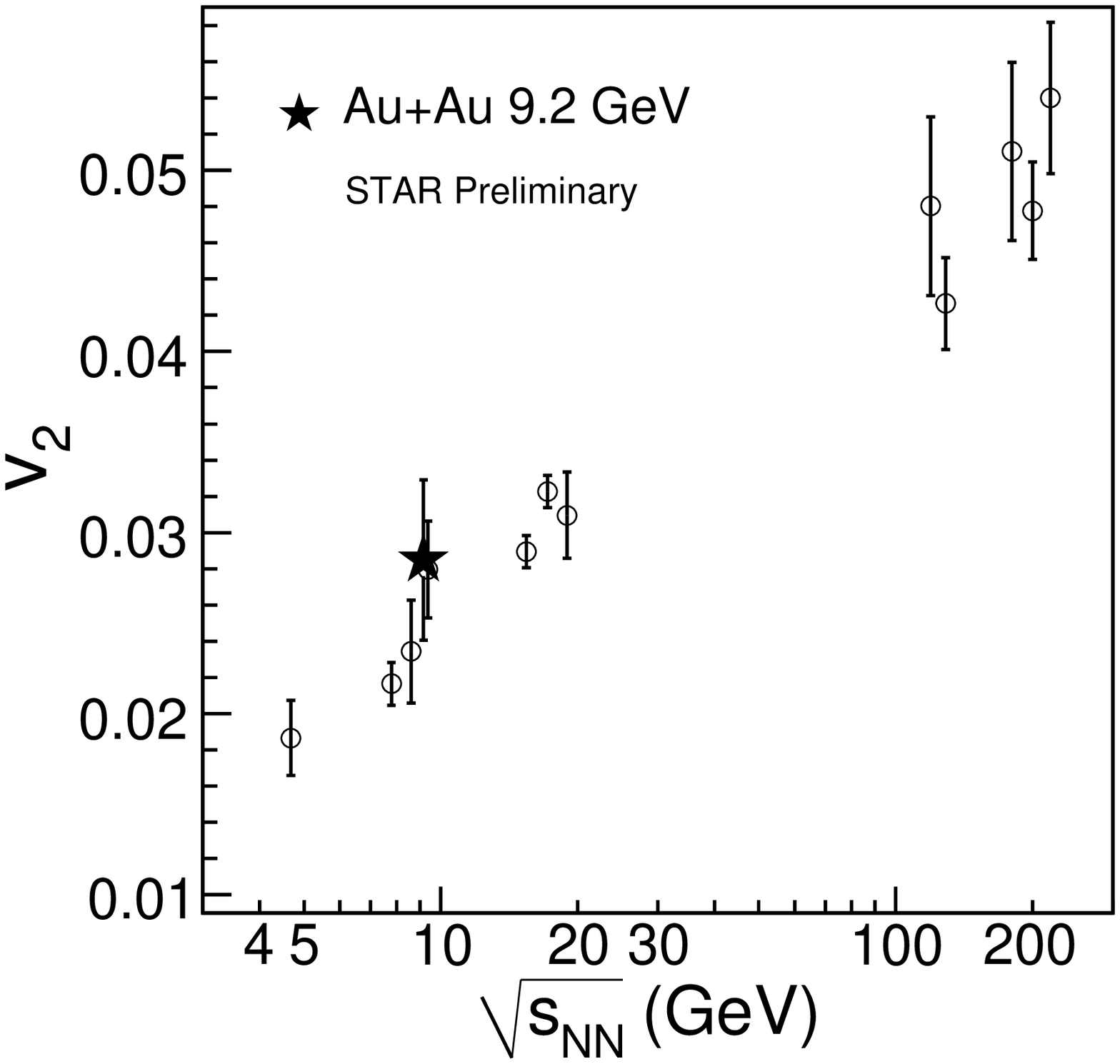}
\vspace{-0.5cm}
\caption{\label{fig:epsart3} Left panel : Charged hadrons $v_{1}$ vs. $\eta$
from 0--60\% 
Au+Au collisions at 9.2 GeV 
(errors shown are statistical). See text for details.
Right panel : Energy dependence of $v_{2}$ near
mid-rapidity ($-1 < \eta < 1$). Errors
are statistical only. See text for details.
}
\end{center}
\end{figure}

Figure~\ref{fig:epsart} shows the particle ratios, $\pi^{-}$/$\pi^{+}$, $\bar{p}/p$,
$K^{-}$/$K^{+}$ and $K$/$\pi$, plotted as a function of $\sqrt{s_{NN}}$. 
Results from 0--10\% central Au+Au collisions at 9.2 GeV (solid stars)
at midrapidity ($|y| < 0.5$) are 
compared with those from AGS~\cite{ags}, SPS~\cite{sps} and RHIC~\cite{STARPID} (open symbols).
We observe that the ratios at 
9.2 GeV are consistent with the $\sqrt{s_{NN}}$ dependence trend. The 
ratio $\pi^{-}$/$\pi^{+}$ (Fig.~\ref{fig:epsart}(a) left panel) for 
9.2 GeV 
%is close to 1.0, 
is close to one, 
suggesting that $\pi^{-}$ and $\pi^{+}$ are produced from similar sources at this energy. 
However,
%this ratio is greater than 1.0 at lower energies, indicating that pions are dominantly produced from
this ratio is greater than one at lower energies, indicating that pions are dominantly produced from
resonance decays (such as $\Delta$). % at lower energies.
The ratio $\bar{p}/p$ (Fig.~\ref{fig:epsart}(b) left panel) at 
9.2 GeV is 
%%%% check %%%%%
%much less than 1, 
much less than one, 
%indicating large baryon stopping due to a large net proton ($p$ - $\bar{p}$) and 
indicating large net proton ($p - \bar{p}$) due to a large baryon stopping and 
hence large value of $\mu_{B}$ at this energy.
%%%% check %%%%%
This ratio increases with increasing center of mass energy,
%approaching a value close to 1 at higher energies. This suggests that $p$ and $\bar{p}$
approaching a value close to one at higher energies. This suggests that $p$ and $\bar{p}$
are dominantly produced by pair production at higher energies. 
%%%%%%%%%%% Mentioning feed down correction %%%%%%%%%%%
The STAR proton results presented here are not corrected 
for feed down contributions.
%%%%%%%%%%% Mentioning feed down correction %%%%%%%%%%%
The 
ratio $K^{-}$/$K^{+}$ (Fig.~\ref{fig:epsart}(a) right panel) for 
9.2 GeV 
is close
to 0.4, which indicates that $\sim$60\% of $K^{+}$ are produced via associated production
%with $\Lambda$. As the center of mass energy increases, this ratio approaches the value of 1, 
with $\Lambda$. As the center of mass energy increases, this ratio approaches the value of one, 
suggesting the dominance of pair production. % mechanism.  
Strangeness production in heavy-ion collision experiments can be studied from the 
kaon to pion ratio (Fig.~\ref{fig:epsart}(b) right panel). A ``horn-like'' distribution is
observed for the $K^{+}$/$\pi^{+}$ ratio around $\sqrt{s_{NN}}$ $\sim$7.7 GeV at the SPS~\cite{sps}. 
It can be seen that 9.2 GeV result %is similar to
agrees with the corresponding results %at SPS.
from SPS data.
This is of great interest to both theorists and experimentalists in order to understand the relevant 
degrees of freedom.

\subsection{Azimuthal anisotropy measurements}% and $\pi$ interferometry}
%Study of Azimuthal anisotropy parameters - directed flow ($v_{1}$) and elliptic flow ($v_{2}$), 
The azimuthal anisotropy parameters - directed flow ($v_{1}$) and elliptic flow ($v_{2}$), 
in ultra-relativistic heavy ion collisions are believed to be sensitive to the equation of state.
Figure~\ref{fig:epsart3} (left panel) shows the charged hadrons $v_{1}$ as a function of 
pseudorapidity ($\eta$) for 0--60\% central Au+Au collisions at 9.2 GeV. The results are compared 
to $v_{1}$ for 30--60\% central Au+Au collisions at 62.4 and 200 GeV~\cite{v14systempaper}. 
Also shown for comparison are $v_{1}$ for charged pions for 0--60\% central Pb+Pb collisions 
at 8.8 GeV~\cite{v2NA49}. The $v_{1}$ for 9.2 GeV shows similar behaviour to that of the other 
center of mass energies at midrapidity. The difference seen at forward rapidities ($|\eta| >$ 2) 
is due to the contributions of spectator protons and nuclear fragments. %to $v_{1}$. 
When $v_{1}$ is divided by the beam rapidities (2.3, 4.2 and 5.4 for $\sqrt{s_{NN}}$ = 9.2 GeV, 
62.4 GeV and 200 GeV, respectively), the difference disappears and $v_{1}$ 
for all $\sqrt{s_{NN}}$ lie on a common trend. Figure~\ref{fig:epsart3} (right panel) shows the
$v_{2}$ as a function of $\sqrt{s_{NN}}$ for charged hadrons. Results from minimum bias collisions 
at 9.2 GeV (solid star symbol) at midrapidity are compared with those from STAR~\cite{flow6_STAR} 
at higher energy, E877~\cite{E877}, NA49~\cite{v2NA49}, PHENIX~\cite{PHENIX} and
PHOBOS~\cite{flow2_PH0BOS} (open circles). The 9.2 GeV $v_{2}$ result follows the established 
$\sqrt{s_{NN}}$ trend.

%\subsection{$\pi$ interferometry measurements}
\subsection{Pion interferometry measurements}
%%%%%%%%%%%%%%% HBT Results %%%%%%%%%%%%%%%%%%%%
%The pion interfermometry measurements for 0-30\% central Au+Au 
%collisions at 9.2 GeV provide radii paramters $R_{\rm out}$, $R_{\rm side}$
%and $R_{\rm long}$ having values 4.81 $\pm$ 0.8, 4.41 $\pm$ 0.5 and 5.06 $\pm$ 0.8 $fm$
%respectively. The ratio $R_{\rm out}$/$R_{\rm side}$ is close to 1 which is consistent
%with the established $\sqrt{s_{NN}}$ dependence trend.
The pion interferometry measurements are performed for 0--30\% central Au+Au 
collisions at 9.2 GeV. Table~\ref{table4} shows various parameters
obtained from these measurements. The ratio $R_{\rm out}$/$R_{\rm side}$ is observed 
to be close 
%to 1 and is consistent with the established $\sqrt{s_{NN}}$ dependence trends. 
to one and is consistent with the established $\sqrt{s_{NN}}$ dependence trends. 
%For a first order phase transition, the ratio $R_{\rm out}$/$R_{\rm side}$ 
%is expected to be much greater than 1.
%For the 9.2 GeV data, this ratio is close to 1 which is consistent
%with the established $\sqrt{s_{NN}}$ dependence trend.
%%%%%%%%%%%%%%% HBT Results %%%%%%%%%%%%%%%%%%%%

\begin{table}
\caption{\label{table4}The HBT parameters for 0--30$\%$ central events and $k_{\rm T}$ $=$ [150, 250] MeV/c.}
\begin{center}
\begin{tabular}{c|c|c|c}
$\lambda$&$R_{\rm out}$ (fm) &$R_{\rm side}$ (fm) &$R_{\rm long}$ (fm) \\
\hline
0.6 $\pm$ 0.1  &  4.8 $\pm$ 0.8 & 4.4 $\pm$ 0.5  &  5.1 $\pm$ 0.8  
\end{tabular}
\end{center}
\end{table}

\section{Summary and Outlook}
We have presented results on identified particle ratios and azimuthal anisotropy 
measurements for Au+Au collisions at 9.2 GeV. 
This is the lowest center of mass energy %injected 
at RHIC so far. The ratios of various particles for 
9.2 GeV
at midrapidity are consistent with the previously established $\sqrt{s_{NN}}$
%dependence trends. The $\pi^{-}$/$\pi^{+}$ ratio for 9.2 GeV is close to 1, 
dependence trends. The $\pi^{-}$/$\pi^{+}$ ratio for 9.2 GeV is close to one, 
%$\bar{p}/p$ ratio is much less than 1, 
$\bar{p}/p$ ratio is much less than one, 
and $K^{-}$/$K^{+}$ ratio is close to 0.4. 
The azimuthal anisotropy measurements and $\pi$ interferometry results
for 
9.2 GeV follow the 
established $\sqrt{s_{NN}}$ dependence trends. The results presented here are only from the 
few thousand good events and qualitative improvements on the previous results
at the SPS will be made with higher statistics. Being a collider experiment, the 
STAR detector has many advantages over a fixed target experiment in terms of 
acceptance ($\eta$, $p_{T}$) and particle density per unit area at a fixed $\sqrt{s_{NN}}$. 
In addition to this, the particle identification in STAR is very good and will be further 
improved by the inclusion of Time Of Flight (TOF). Based on the results presented here
for Au+Au collisions at
9.2 GeV and capabilities of STAR, % experiment, we can say 
it is clear that the RHIC collider and the STAR experiment are ready for the 
future ``Critical Point Search'' program.
%%%%%%%%%%%%%% copied %%%%%%%%%%%%%%%%%%%

%% end of main text

%\section*{Acknowledgments} % please check/modify, comment out or delete if not needed
%This is where one places acknowledgments for funding bodies etc., if needed.
%For the large collaborations, this is listed once and for all, together with 
%the author lists etc. in the proceedings back-material.

 % do not change 
\end{document}